# Landau-Ginzburg-Devonshire theory for electromechanical hysteresis loop formation in piezoresponse force microscopy of thin films

A.N. Morozovska,[1,1] E.A. Eliseev,[2] S.L. Bravina[3] and S.V. Kalinin[4]

[1] Institute of Semiconductor Physics, National Academy of Sciences of Ukraine,
41, pr. Nauki, 03028 Kiev, Ukraine

[2] Institute for Problems of Materials Science, National Academy of Sciences of Ukraine,
3, Krjijanovskogo, 03142 Kiev, Ukraine

[3] Institute of Physics, National Academy of Sciences of Ukraine,
46, pr. Nauki, 03028 Kiev, Ukraine

[4] The Center for Nanophase Materials Sciences, Oak Ridge National Laboratory,
Oak Ridge, TN 37922

Electromechanical hysteresis loop formation in piezoresponse force microscopy of thin ferroelectric films is studied with special emphasis on the effects of tip size and film thickness, as well as dependence on the tip voltage frequency. Here, we use a combination of Landau-Ginzburg-Devonshire (LGD) theory for the description of the local polarization reversal with decoupling approximation for the calculation of the local piezoresponse loops shape, coercive voltages and amplitude. LGD approach enables addressing both thermodynamics and kinetics of hysteresis loop formation. In contrast to the "rigid" ferroelectric approximation, this approach allows for the piezoelectric tensor components dependence on the ferroelectric polarization and dielectric permittivity. This model rationalizes the non-classical shape of the dynamic piezoelectric force microscopy (PFM) loops.

---

[1] morozo@i.com.ua

## 1. Introduction

Polarization dynamics in ferroelectric and multiferroic materials underpins a broad range of information technology applications including ferroelectric random access memories [1], ferroelectric gate field-effect transistors [2], and tunneling barriers [3, 4, 5, 6, 7, 8]. Beyond these applications, polarization dynamics is strongly linked to enhanced piezoelectric properties in disordered ferroelectrics [9] and hence is broadly utilized in microelectromechanical systems, sensors, and actuators. Finally, polarization switching in ferroelectrics offers a convenient model phenomenon for more complex electrochemical and thermal phase transformation [10].

In the last decade, piezoresponse force microscopy has emerged as a powerful tool of imaging static and dynamics domains structures in ferroelectric materials and device structures [11, 12, 13, 14, 15, 16]. Beyond imaging, single point hysteresis loop measurements in PFM [17, 18, 19, 20, 21, 22, 23, 24, 25] and spectroscopic hysteresis loop imaging allowed local probing of polarization dynamics on the nanoscale, as summarized in detail in reviews [26, 27]. In particular, capability for probing [28, 29] and manipulating [30] non-180$^o$ switching and also local polarization reversal on a single defect level [16] were demonstrated.

Understanding hysteresis loop shape peculiarities in PFM spectroscopy requires a solution of a number of independent problems, including the kinetics and thermodynamics of domain formation under the biased SPM tip, relationship between domain size and measured PFM signal, and tip calibration. The thermodynamics of domain formation in the tip field has been extensively studies for uniaxial [31, 32, 33, 34, 35, 36, 37, 38] and multiaxial [39] ferroelectrics in the rigid ferroelectric approximation. The kinetics of this process was studied by Molotskii et al [40, 41]. Recently, a number of studies of thermodynamics of domain formation in the phase-field approximation were reported [17, 30, 42, 43, 44]. The relationship between domain size and PFM signal at this point is available only for a rigid ferroelectric model, as studied by Kalinin et al. [45] in linear decoupling theory [46, 47]. Using the decoupling theory Morozovska et al [48, 49] calculated the size effects of the local piezoresponse in thin ferroelectric films.

Finally, a number of studies on calibration of the PFM tip geometry based on observed domain wall width have been reported [50, 51]. In particular, Tian et al [50] studied the dependence of the local piezoresponse of the LNO ferroelectric single crystal for different radiuses of the PFM probe tip. They obtained that effective piezoresponse of the LNO single domain region is independent on the tip radius, while the halfwidth of the 180-degree domain wall measured by PFM and calculated theoretically (both analytically and by finite elements modeling) increases with the tip radius increase both for squashed and spherical tips. Rigorously speaking, the effective piezoresponse is independent on the tip radius until the electric voltage applied to the PFM tip is small enough. For voltages higher than the local coercive bias, local polarization reversal appears and results in the nanodomain formation and further growth (see [26] and Refs. therein).

Here, we analytically study the finite size effects in the dynamic PFM hysteresis loop originating from the interplay between the PFM tip radius and film thickness. This goal is achieved through the combination of analytical Landau-Ginzburg- Devonshire (LGD) theory, for the description of the local polarization reversal with the decoupling approximation for the calculation of the local piezoresponse loop shape, coercive voltages and amplitude. In contrast to the "rigid" approximation, mainly used previously for PFM hysteresis loops calculations [52, 53, 54, 55], LGD approach [42, 43, 44] allows taking into account the dependence of the piezoelectric tensor components on the ferroelectric polarization and dielectric permittivity. LGD approach can describe the nonlinear dependence of the stress piezoelectric tensor components on the applied voltage and "bumps" on PFM hysteresis loop, which are sometimes observed experimentally [56, 57].

### II. Problem statement and basic equations

Here, we analytically study finite size effects of the dynamic PFM hysteresis loop originated from the interplay between the PFM tip radius and film thickness. The spontaneous polarization **P** of ferroelectric film is directed along the polar axis, *z*. The sample is dielectrically isotropic in transverse directions, i.e. permittivities $\varepsilon_{11} = \varepsilon_{22}$, while $\varepsilon_{33}$ value may be different (see **Fig. 1**). Coordinate system $\mathbf{r} = (\xi_1, \xi_2, z)$ is linked to the probe apex, coordinates $\mathbf{y} = (y_1, y_2, z)$ indicate the tip apex position in the sample coordinate system **y**. *V* is the voltage applied to the PFM tip.

In the case of effective point charge model, the probe apex electric field can be modeled by a single charge $Q = 2\pi\varepsilon_0\varepsilon_e R_0 V(\kappa+\varepsilon_e)/\kappa$ located at distance $d = \varepsilon_e R_0/\kappa$ for a spherical electrode with curvature $R_0$ ($\kappa = \sqrt{\varepsilon_{11}\varepsilon_{33}}$ is the effective dielectric constant determined by the dielectric permittivity in z-direction, $\varepsilon_e$ is the ambient dielectric constant), or $d = 2R_0/\pi$ for a squashed tip electrode represented by a disk of radius $R_0$ in contact with the sample surface [52]. The contribution of the probe conical part is negligible in the vicinity of the tip-surface contact junction, justifying the use of the effective point charge approximation with high accuracy [58].

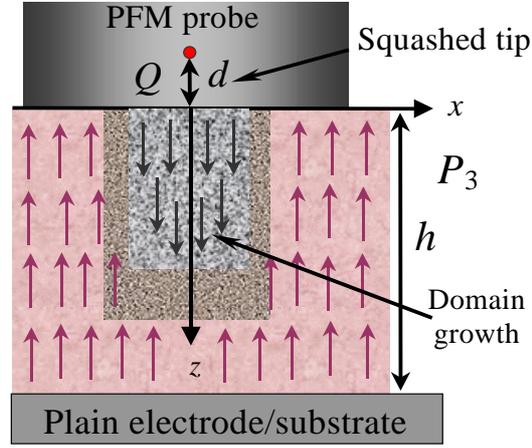

**FIG. 1**. Nanodomain caused by the electric field of the charged PFM probe in contact with the ferroelectric film surface

For most inorganic ferroelectrics, the elastic properties are weakly dependent on orientation. Hence, material can be approximated as elastically isotropic. For the sample homogeneous in the several penetration depths of electric field, vertical surface displacement below the tip, i.e. vertical PFM signal $u_3(\mathbf{0},\mathbf{y})$, can be calculated in the decoupling approximation [45, 48-49] as:

$$u_3(\mathbf{0},\mathbf{y}) = \int_{-\infty}^{\infty} d_{mnk}(\mathbf{y}-\boldsymbol{\xi})\left(\int_0^h c_{jlmn} E_k(-\xi_1,-\xi_2,z)\frac{\partial}{\partial \xi_l} G_{3j}(\xi_1,\xi_2,z)dz\right)d\xi_1 d\xi_2. \quad (1)$$

Here $d_{ijk}(\mathbf{r}) = \dfrac{\partial u_{ij}(\mathbf{r})}{\partial E_k(\mathbf{r})} = 2\varepsilon_0 Q_{lmjk}\varepsilon_{il}(\mathbf{r})P_m(\mathbf{r})$ are the stress piezoelectric tensor components ($u_{ij}$ is the strain tensor), $Q_{lmjk}$ is electrostriction tensor for cubic symmetry, $c_{kjmn}$ are stiffness tensor components. For considered uniaxial ferroelectric polarization **the** nonzero piezoelectric coefficients are $d_{33} = 2\varepsilon_0 Q_{11}\varepsilon_{33}P_3$, $d_{31} = 2\varepsilon_0 Q_{12}\varepsilon_{33}P_3$ and $d_{15} = 2\varepsilon_0 Q_{44}\varepsilon_{11}P_3$ in Vogt notation. The Green's tensor $G_{ij}(\xi)$ for elastically isotropic films is calculated in Refs.[48, 49]. Note, that using the dependence of piezoelectric coefficient on dielectric permittivity, allows the decoupling approximation to be extended for "soft" ferroelectrics with pronounced coordinate and field dependence of the dielectric permittivity. This approach is especially valuable given that electrostriction tensor components are generally field- and temperature independent. At the same time, the latter can be readily determined form the LGD theory for arbitrary thermal field, and mechanical conditions [59], as is analyzed below.

The PFM tip generates the electric field $E_k(\mathbf{r})$ at the point $\mathbf{r} = (x, y, z)$ within the sample. In the case of effective point charge model, the spatial distribution of the electric field z-component produced by the charged PFM probe under the surface of ferroelectric film of thickness $h$ can be approximated as [42]:

$$E_3(\rho, z, t) = V(t)\int_0^\infty dk J_0(k\rho) \cdot \frac{\cosh(k(h-z)/\gamma)}{\sinh(kh/\gamma)}\frac{kd}{\gamma}\exp(-kd)$$

$$\approx V(t)\begin{cases} \dfrac{\gamma d}{h^2}\left(\dfrac{1}{2}\psi^{(1)}\left(\dfrac{\gamma d}{2h}\right) - \left(\dfrac{h}{\gamma d}\right)^2\right), & z = 0, \ \rho < d, \\ \dfrac{\gamma d}{2h^2}\psi^{(1)}\left(\dfrac{h + \gamma d}{2h}\right), & z = h, \ \rho < d. \end{cases} \quad (2)$$

Here the polar radius $\rho = \sqrt{x^2 + y^2}$, PolyGamma function $\psi^{(1)}(x)$ is the first derivative of the digamma function $\psi(x) = \Gamma'(x)/\Gamma(x)$, and $\Gamma(x)$ is the gamma function. The time-dependent voltage $V(t)$ applied to the tip represents both the ac and dc components of tip bias, $\gamma = \sqrt{\varepsilon_{33}/\varepsilon_{11}}$ is the dielectric anisotropy factor, $d$ is the effective distance determined by the tip electrode - film geometry (see Refs.[49] and **Fig. 1**).

Note, that we consider constant dielectric permittivity ($\varepsilon_{ij}$ = const) when deriving Eq. (2). Rigorously speaking, the permittivity should be coordinate and voltage dependent allowing for the distribution of ferroelectric polarization, e.g. $\varepsilon_{33}(\mathbf{r}) = \varepsilon_{33}^b + \frac{1}{\varepsilon_0}\frac{\partial P_3(\mathbf{r})}{\partial E_3}$ ($\varepsilon_{33}^b$ is the background permittivity [60]). In Eq.(2) we neglect the coordinate dependence $\varepsilon_{ij}(\mathbf{r})$ but consider the electric field dependence of the averaged value $\varepsilon_{ij} = \langle \varepsilon_{ij}(\mathbf{r}) \rangle$. Consequently the permittivity field dependences should be taken into account in piezoelectric stress tensor coefficients $\langle d_{ijk}(\mathbf{r}) \rangle \approx 2\varepsilon_0 Q_{lmjk} \langle \varepsilon_{il}(\mathbf{r}) \rangle P_m(\mathbf{r})$. The role of permittivity is analyzed in Section **IV.3.**

In the continuous media approximation both polarization **P** and its second normal derivative Δ**P** are small in the immediate vicinity of domain wall boundary, since the boundary is axially symmetric and both **P** and Δ**P** are identically zero for an uncharged planar boundary between 180-degree domains. Δ**P** becomes non-zero only if boundary is curved and charged, but for the examined case the deviation is as small as the ratio of the domain wall thickness $L_\perp$ (several lattice constants) to the tip effective size $d$ [42]. Under the typical condition of rapid domain intergrowth ("domain breakdown") through the depth of thin ferroelectric film the charged domain wall disappears and so depolarization field vanishes (walls for cylindrical domain are 180-degree).

Under the typical condition *of thin domain walls,* $L_\perp \ll d$, a thermodynamically stable domain wall boundary ρ(z) can be estimated from the Eq.(2) under the condition $E_3(\rho,z) = E_c$, where $E_c$ is the coercive field and $E_3(\rho,z)$ is given by Eq.(1). The thermodynamic coercive field $E_c$ of the ferroelectric film is determined as [61]:

$$E_c = \frac{2}{3\sqrt{3}}\sqrt{-\frac{\alpha^3}{\beta}}, \qquad (3a)$$

$$E_c = \frac{2}{5}\left(2\beta + \sqrt{9\beta^2 - 20\alpha\delta}\right)\left(\frac{2\alpha}{-3\beta - \sqrt{9\beta^2 - 20\alpha\delta}}\right)^{3/2}. \qquad (3b)$$

Here Eq.(3a) is valid for second order ferroelectrics, and Eq.(3b) is valid for the first order ferroelectrics respectively. The LGD free energy expansion coefficient $\delta > 0$, while $\beta < 0$ ($\beta > 0$) for first (second) order phase transitions. The coefficient $\alpha < 0$ in the ferroelectric phase. The coefficient $\alpha(h, R_0)$ is renormalized by the finite size effects, screening conditions and elastic stress in thin films [62, 63, 64].

### III. Finite size effect on coercive voltage
### III.1 Coercive voltage in LGD local approach

The local approach allows the functional dependence of the coercive voltage $V_c$ on the effective tip size $d$ and the film thickness $h$ from the algebraic equation of domain nucleation $E_3(\rho, z) = E_c$ to be estimated as:

$$V_c^{loc}(z) = \begin{cases} E_c \dfrac{h^2}{\gamma d}\left(\dfrac{1}{2}\psi^{(1)}\!\left(\dfrac{\gamma d}{2h}\right) - \left(\dfrac{h}{\gamma d}\right)^2\right)^{-1}, & z=0, \rho=0, \\[2ex] \dfrac{2h^2 E_c}{\psi^{(1)}\!\left(\dfrac{h+\gamma d}{2h}\right)\gamma d} \underset{h\gg\gamma d}{\approx} \dfrac{\gamma E_c}{d}\left(d+\dfrac{h}{\gamma}\right)^2, & z=h, \rho=0. \end{cases} \qquad (4)$$

Dependences of $V_c^{loc}(0)$ and $V_c^{loc}(h)$ on the ratio $\gamma d/h$, proportional to the effective tip size $d$, are shown in **Fig. 2a,b** (compare the dependences with the dependences $V_c^{loc} \sim d$ shown in Fig. 3 from Ref.[42]).

Note, that the voltage averaged over the cylindrical region $\{d, h\}$ proved to be independent on the tip effective size: $\langle V_c(z,\rho)\rangle_{r\sim d} \sim h\cdot E_c$. The result agrees with the model introduced in Refs. [65, 66], where the electric field was integrated over the actual region of domain nucleation.

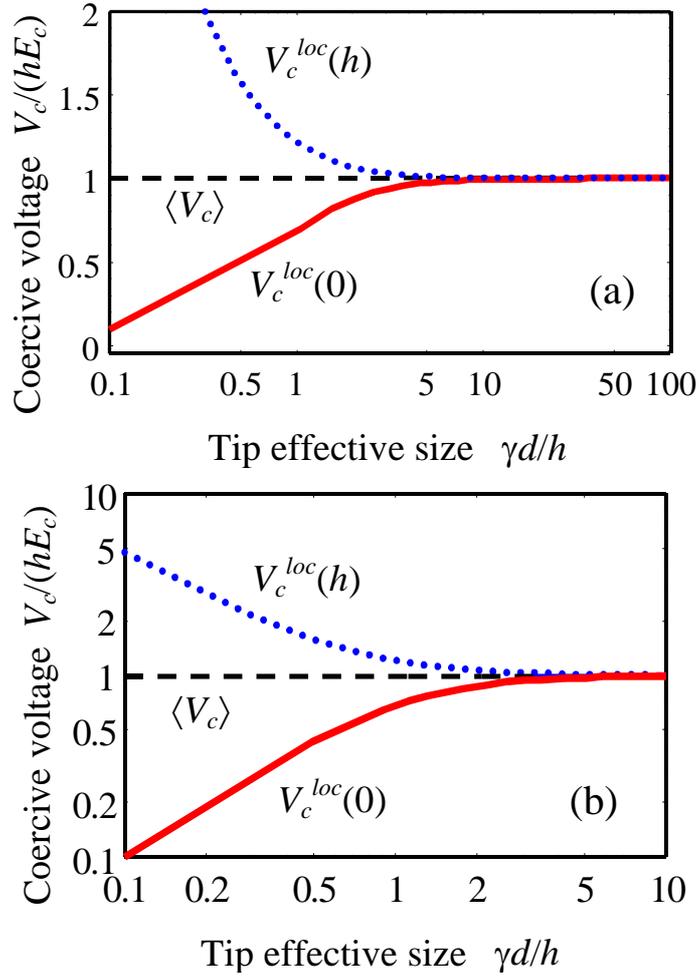

**FIG. 2**. (a) Dependence of the static coercive voltage $V_c^{loc}(0)$, $V_c^{loc}(h)$ and $V_c^{eff}$ on the dimensionless tip size $\gamma d/h$ (log-linear scale). Plot (b) shows the same dependences in the log-log scale.

### III.2 Coercive voltage in Burtsev and Chervonobrodov approach

One of the significant factors affecting polarization switching is the discreetness of the atomic lattice, for which periodic potential offers a pinning landscape for the domain wall. This effect was originally considered by Miller and Weinreich [67] for the idealized discrete lattice model. Burtsev and Chervonobrodov [68] considered a more realistic model with continuous lattice potential and diffuse domain walls, at that the nucleus shape and domain wall width are calculated self-consistently. This approach was recently developed by Rappe for the atomistic case [69].

Using Burtsev and Chervonobrodov approach we obtained the equation for the coercive bias [44]:

$$k_B T = F_a^{BC} = \sqrt{\ln\left(\frac{\sqrt{\sigma_{min}\delta\sigma}}{2cP_S E_c}\right)\frac{\left(c\sqrt{\sigma_{min}\delta\sigma}\right)^3}{4\pi\varepsilon_0\varepsilon_{11}}} \frac{1}{E_c}, \quad (5a)$$

where

$$E_c = V_c \frac{\gamma d}{h^2}\left(\frac{1}{2}\psi^{(1)}\left(\frac{\gamma d}{2h}\right) - \left(\frac{h}{\gamma d}\right)^2\right). \quad (5b)$$

At $h \gg \gamma d$ the approximation $E_c \sim \frac{V_c}{d}$ is valid. Since the logarithm is the slowly varying function, the coercive field could be estimated from the condition

$$E_c = \sqrt{\ln\left(\frac{\sqrt{\sigma_{min}\delta\sigma}}{2cP_S E_c}\right)\frac{\left(c\sqrt{\sigma_{min}\delta\sigma}\right)^3}{4\pi\varepsilon_0\varepsilon_{11}}} \frac{1}{k_B T} \approx \sqrt{\frac{\left(c\sqrt{\sigma_{min}\delta\sigma}\right)^3}{4\pi\varepsilon_0\varepsilon_{11}}} \frac{1}{k_B T}. \quad (5c)$$

Thus one again leads to the dependence (4) for $V_c^{loc}(0)$.

To summarize the results of Subsections III.1 and III.2, the expression

$$V_c(d,h) \approx E_c \frac{h^2}{\gamma d}\left(\frac{1}{2}\psi^{(1)}\left(\frac{\gamma d}{2h}\right) - \left(\frac{h}{\gamma d}\right)^2\right)^{-1} \quad (6)$$

should be regarded as the most relevant one for the estimation of domain nucleation voltage and can be used for estimation of corresponding size effects.

The voltage dependence of the domain radius can then be estimated as (see Table 1 in Ref.[42] and Eq.(5)):

$$r(V) \approx \begin{cases} d\sqrt{\left(\frac{\gamma d V}{E_c h^2}\right)^{2/3}\left(\frac{1}{2}\psi^{(1)}\left(\frac{\gamma d}{2h}\right) - \left(\frac{h}{\gamma d}\right)^2\right)^{2/3} - 1}, & h \gg \gamma d, \\ d\sqrt{\left(\frac{\gamma d V}{E_c h^2}\right)^{2}\left(\frac{1}{2}\psi^{(1)}\left(\frac{\gamma d}{2h}\right) - \left(\frac{h}{\gamma d}\right)^2\right)^{2} - 1}, & h \ll \gamma d. \end{cases} \quad (7)$$

Note, that this analysis essentially justifies early arguments of Kolosov [70] stating that the domain size in a PFM experiment corresponds to the region in which tip-induced field exceeds the coercive field. Morozovska et al [42] obtained a similar result for semi-infinite ferroelectrics, but provided that the sum of the nascent domain depolarization field and the **probe** tip-induced field exceed**s** the coercive field, at that the positive depolarization field in front of the moving counter domain wall is the driving force of the observed domain tip elongation in the region where the probe electric field is much smaller than the intrinsic coercive field. Moreover, for an infinitely thin domain wall the depolarization field outside the semi-ellipsoidal domain tip is always higher than the intrinsic coercive field, that must initiate the local domain breakdown through the film depth, while the domain length is finite in the energetic approach evolved by Landauer [31] and Molotskii [31].

## IV. Electromechanical response
### IV.1. Response in decoupled approximation

The measured parameter in a PFM experiment is the electromechanical response of the probe as a function of slowly varying dc bias [26, 27]. Using decoupling theory, the vertical local piezoresponse from a cylindrical domain of radius $r(V)$ can be estimated as:

$$d_{33}^{eff}(V,t) \approx -2\varepsilon_0 P_V(t)\left(\varepsilon_{33}(Q_{11}w_{333} + Q_{12}w_{313}) + \varepsilon_{11}Q_{44}w_{351}\right). \quad (8)$$

The maximal polarization $P_V(t)$ satisfies the "local" LGD-equation:

$$\left(\Gamma\frac{d}{dt} + \alpha + \beta P_V^2 + \delta P_V^4\right)P_V \approx V(t)\frac{\gamma d}{h^2}\left(\frac{1}{2}\psi^{(1)}\left(\frac{\gamma d}{2h}\right) - \left(\frac{h}{\gamma d}\right)^2\right). \quad (9)$$

Numerical comparison of the Eq.(1) with approximate Eq.(9) proves that the discrepancy is small and the accuracy about 5-10% is reached for a typical range of parameters.

Functions $w_{3jk}$ are the normalized object transfer function components [45], which Pade approximations we derived as:

$$w_{333} \approx \left(\frac{\kappa(\varepsilon_e + \kappa_b)}{\kappa_b(\varepsilon_e + \kappa)}\frac{\gamma d}{h} + \frac{(1+\gamma)^2}{(1+2\gamma)}\right)^{-1}\left(1 - \left(1 + \frac{\kappa_b + \varepsilon_e}{\kappa - \kappa_b} + \frac{h}{\gamma d}\frac{\varepsilon_e - \kappa}{\kappa\ln(1-\chi)}\right)^{-1}\right), \quad (10a)$$

$$w_{351} \approx \left(\frac{(\varepsilon_e + \kappa_b)}{(\varepsilon_e + \kappa)}\frac{(1-\nu)}{\nu}\frac{d^2}{h^2} + \frac{(1+\gamma)^2}{\gamma^2}\right)^{-1}\left(1 - \left(1 + \frac{\kappa_b + \varepsilon_e}{\kappa - \kappa_b} + \frac{h}{\gamma d}\frac{\varepsilon_e - \kappa}{\kappa\ln(1-\chi)}\right)^{-1}\right), \quad (10b)$$

$$w_{313} \approx \left(\frac{(1-\nu)}{2\nu}\frac{\kappa(\varepsilon_e + \kappa_b)}{\kappa_b(\varepsilon_e + \kappa)}\frac{\gamma d}{h} + \frac{(1+\gamma)^2}{1+2\nu(1+\gamma)}\right)^{-1}\left(1 - \left(1 + \frac{\kappa_b + \varepsilon_e}{\kappa - \kappa_b} + \frac{h}{\gamma d}\frac{\varepsilon_e - \kappa}{\kappa\ln(1-\chi)}\right)^{-1}\right) \quad (10c)$$

Here ν ~ 1/3 is the Poisson ratio. The parameter

$$\chi = \left(\frac{\kappa_b - \kappa}{\kappa_b + \kappa}\right)\left(\frac{\varepsilon_e - \kappa}{\varepsilon_e + \kappa}\right). \tag{11}$$

Always $|\chi| \leq 1$, $\varepsilon_e$ is the dielectric constant of the ambient, $\kappa = \sqrt{\varepsilon_{33}\varepsilon_{11}}$ is **the** effective dielectric constant, $\gamma = \sqrt{\varepsilon_{33}/\varepsilon_{11}}$ is the dielectric anisotropy factor of the film (or surface layer), $\kappa_b = \sqrt{\varepsilon_{33}^b \varepsilon_{11}^b}$ is effective dielectric constant of the substrate or bottom electrode. Exact series for $w_{3jk}$ are listed in Ref.[49].

Local piezoresponse hysteresis loops are shown in the **Fig. 3a,b**. Note that the loop features (coercive voltage $V_c$ and maximal amplitude $u_{max}$) depend on the ratio $\gamma d/h$ as well as on the frequency ω of the voltage $V = V_0 \sin(\omega t)$ applied to the PFM tip. The hysteresis loop width monotonically increases $\gamma d/h$ with the frequency ω decrease. Actually, smaller tip sizes correspond to the higher tip electric field below the film surface. The hysteresis loops blow with the dimensionless frequency $w = \omega\Gamma/|\alpha|$ increase (compare plots a and b).

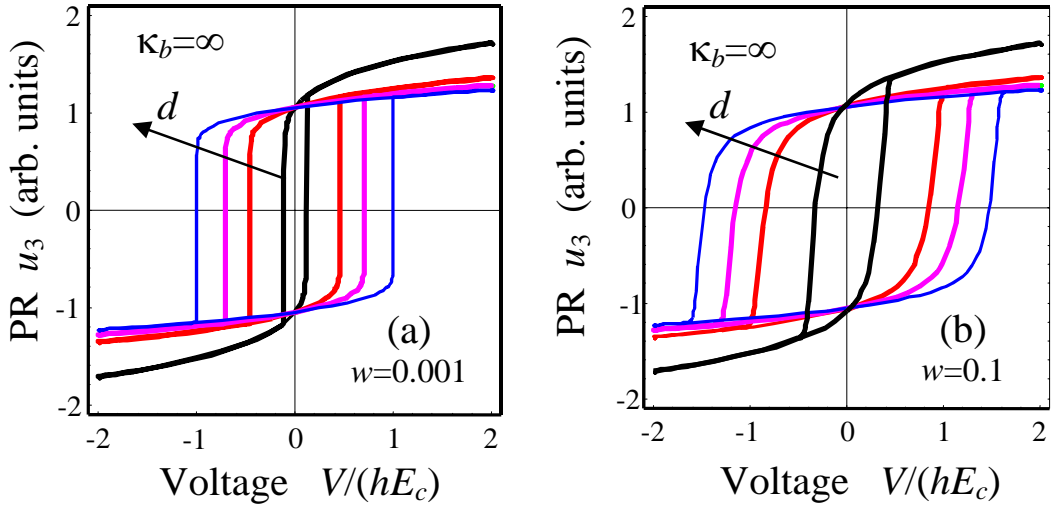

**FIG. 3**. Local piezoresponse (PR) hysteresis loops calculated for different tip effective sizes $\gamma d/h = 0.1, 0.5, 1, 5$ (d increases in the arrow direction), frequency $w = 0.001$ (a), 0.1 (b) and $\kappa_b \to \infty$. All plots are generated for the ferroelectric film with the first order phase transition. Parameter $\beta/\sqrt{-\alpha\delta} = -0.2$.

## IV.2. Dynamic effects on loop shape

The dynamic size effects, i.e. the dependence of $V_c(\omega)$, maximal PFM amplitude $u_3(V_{max})$ and remnant displacement $u_3(V=0)$ on the ratio $\gamma d/h$, are shown in **Figs. 4a,c** for several values of dimensionless frequency $w = \omega\Gamma/|\alpha|$.

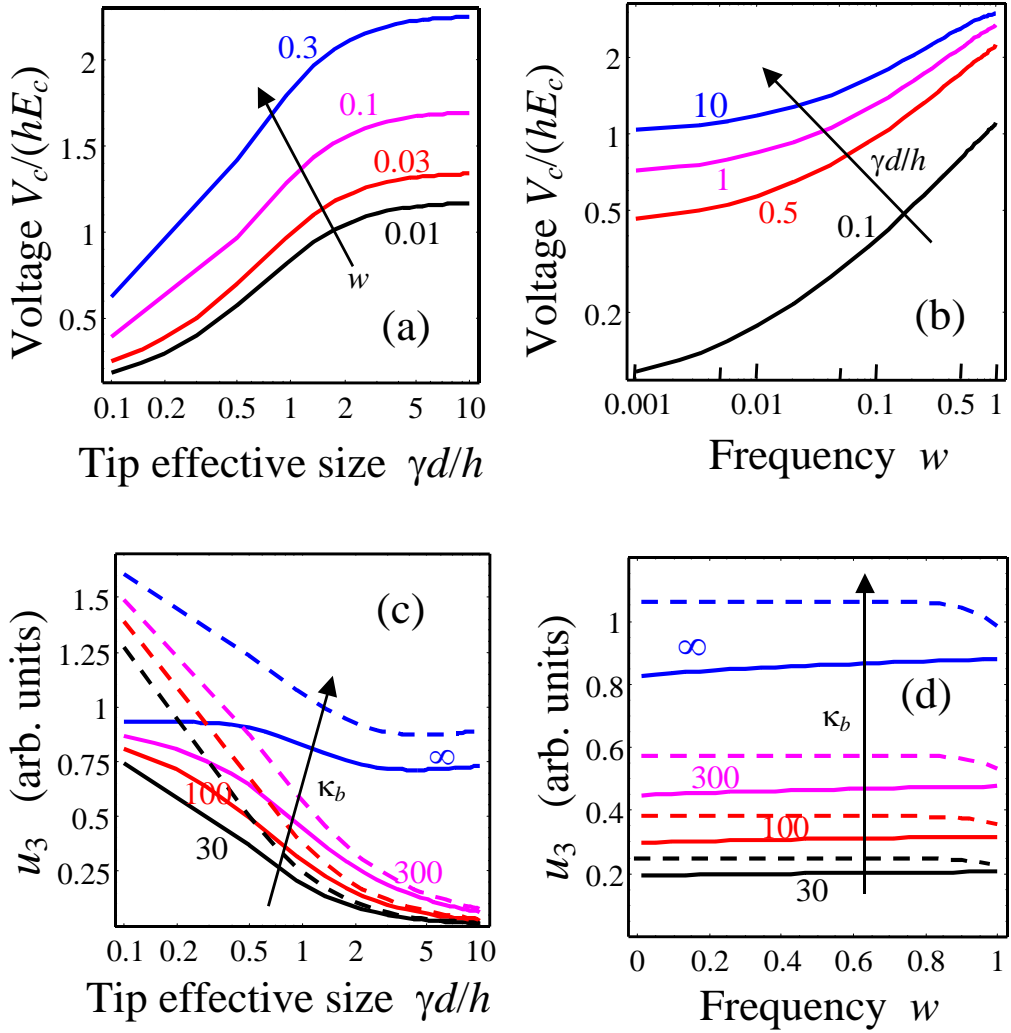

**FIG. 4**. (a) Dynamic coercive voltage vs. the ratio $\gamma d/h$ calculated for **the** different frequencies $w = 0.01, 0.03, 0.1, 0.3$ ($w$ increases in the arrow direction). (b) Dynamic coercive voltage vs. the frequency $w$ for the different ratios $\gamma d/h = 0.1, 0.5, 1, 10$ (increases in the arrow direction). (c) Maximal PFM amplitude $u_3(V_{max})$ (dashed curves, $V_{max} = 3V_c$) and remnant displacement $u_3(V=0)$ (solid curves) vs. the ratio $\gamma d/h$ calculated for frequency $w = 0.03$ and different substrate permittivities $\kappa_b = \infty$ (metal), 300 (SrTiO$_3$), 100 (TiO$_2$), 30 (label near the curves). (d) $u_3(V_{max})$ (dashed curves, $V_{max} = 3V_c$) and $u_3(V=0)$ (solid curves) vs. the frequency $w$ for different $\kappa_b = \infty$, 300, 100, 30 (label near the curves). All plots are generated for $\beta/\sqrt{-\alpha\delta} = -0.2$.

It is seen from **Fig. 4a** that the coercive voltage monotonically increases with $\gamma d/h$ increase and eventually saturates at $\gamma d/h \gg 1$. Similarly, from **Fig. 4b** that the coercive voltage increases with $w$ increase. Maximal amplitude $u_3(V_{max})$ and remnant displacement $u_3(V=0)$ monotonically decreases with $\gamma d/h$ increase (**Fig. 4c**). Coercive voltage is independent on $\kappa_b$ value, while $u_3$ decreases with $\kappa_b$ decrease (compare different curves in **Figs. 4c,d**). Note, that that the limit $\kappa_b \to \infty$ corresponds to the conducting substrate or metallic electrode, when the extrinsic contribution to the size effect is negligible allowing for the electric field homogeneity in the ultrathin ferroelectric films near the substrate [48]. Frequency dispersion is noticeable for the coercive voltage (see discrepancies between the curves for different $w$ in **Figs. 4a** and **Figs. 4b**) and almost absent for the displacement amplitude (see **Figs. 4d**).

### IV.3. Role of dielectric permittivity on the loop shape

Dielectric permittivity $\varepsilon_{33}$ was regarded as voltage independent constant in **Figs. 2-4.** Consequently, the shape of the PFM hysteresis loop reproduces the shape of the ferroelectric polarization one, which is in a reasonable agreement with multiple experiments (see [27] and refs therein). However, generally dielectric constant in ferroelectric can be a strong function of electric field. Here, we incorporate the nonlinearity effects on PFM loops, i.e. the voltage dependent dielectric permittivity as $\varepsilon_{33}(V) = \varepsilon_{33}^b + \varepsilon_0^{-1}\langle \partial P_3/\partial E_3 \rangle$, where averaging is performed over the entire cylindrical region of domain formation.

LGD equations for an average dielectric permittivity $\langle \chi_{33} \rangle = \langle \partial P_3 / \partial E_3 \rangle$ acquire the form:

$$\left( \Gamma \frac{d}{dt} + \alpha_S + 3\beta P_S^2 + 5\delta P_S^4 \right) \langle \chi_{33} \rangle = 1, \qquad (12a)$$

here the polarization $P_S$ is determined from the equation

$$\left( \Gamma \frac{d}{dt} + \alpha_S \right) P_S + \beta P_S^3 + \delta P_S^5 = \langle E_3(\rho, z) \rangle. \qquad (12b)$$

The field $\langle E_3(\rho, z) \rangle$ is averaged over the domain volume (see **Appendix A**). The coefficient $\alpha_S$ can be different from coefficient $\alpha$ in Eq.(9) allowing for the motion of multiple domain walls and well as the polarization gradient effect [71].

Results of our calculations are shown in **Fig. 5.** It is seen from the **Fig. 5c**, that the bumps on the piezoresponse loops originated from the maximums of the dielectric susceptibility $\varepsilon_{33}(V)$ near its coercive bias (compare the maximum positions in **Figs. 5b** and **5c**). Note, that coercive bias is the same for ferroelectric and PFM hysteresis loops, but not for the permittivity ones (compare vertical plots **a, b, c** and **d, e, f**).

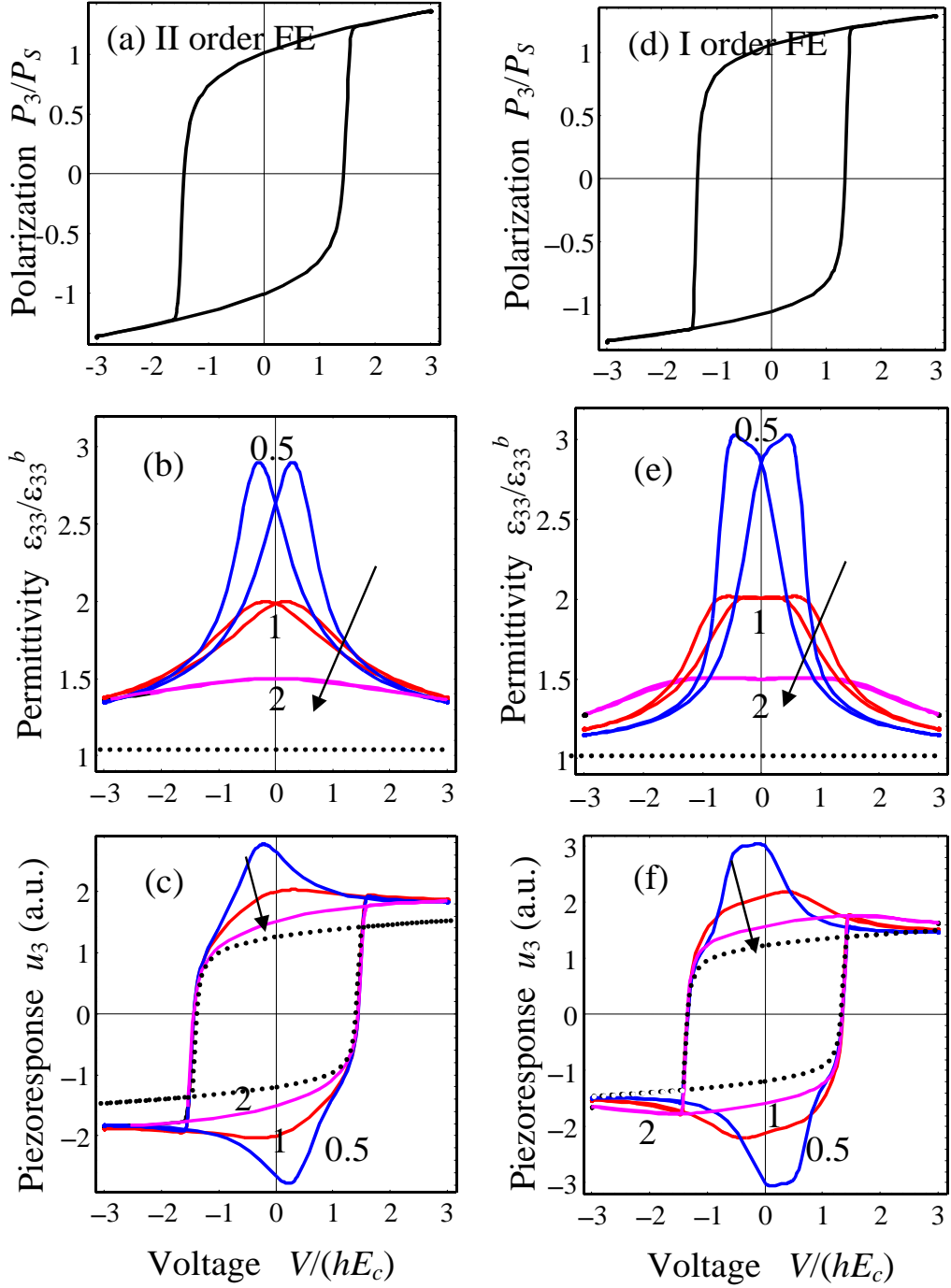

**FIG. 5**. (a,d) Polarization, (b,e) dielectric permittivity and (c,f) local piezoresponse loops calculated for tip effective size $\gamma d/h = 10$, frequency $w = 0.03$, and $\kappa_b \to \infty$. Different permittivity and piezoresponse loops correspond to the different $\alpha_S/(-\alpha) = 2, 1, 0.5$ (figures near the curves) and $\beta/\sqrt{-\alpha\delta} = -0.2$ for the first order ferroelectrics (II order FE –left

column, II order FE – right column). Dotted loops are plotted for the voltage independent permittivity $\varepsilon_{33}$. Coefficient $\alpha_S$ increases in the arrow direction.

**Summary**

The size and dynamic effects on the hysteresis loop**s** in the piezoresponse force microscopy of thin films are explored using the combination of Landau-Ginzburg-Devonshire theory and decoupled approximation. The local piezoresponse loops shape, coercive voltages and amplitude dynamic behaviors and scaling with film thickness are determined. In contrast to the "rigid" approximation, mainly used previously for PFM hysteresis loops calculations, we take into account the explicit dependence of the piezoelectric tensor components on the ferroelectric polarization and dielectric permittivity. It is shown that dielectric nonlinearity can lead to the formation of the non-monotonic PFM hysteresis loops.

Overall, the developed framework enables calculation of PFM responses and hysteresis loops for arbitrarily complex geometries using e.g. experimental or numerically calculated domain configurations, and can also be extended to more complex electromechanical phenomena in e.g. electrochemical systems.

**Acknowledgements**

Authors are grateful to Prof. N.V. Morozovsky for useful discussions and critical remarks. Research for ANM and EAE is sponsored by the Ministry of Education and Sciences of Ukraine (Grant UU30/004, joint with National Science Foundation DMR-0908718 and DMR-0820404). The work is supported in part (SVK) by the Scientific User Facilities Division, US DOE.

**Appendix A**

In the framework of the LGD phenomenology, a stable or metastable spontaneous polarization distribution inside the proper ferroelectric can be found as the solution of the stationary LGD equation:

$$\alpha P_3 + \beta P_3^3 + \delta P_3^5 - \xi \frac{\partial^2 P_3}{\partial z^2} - \frac{\eta}{\rho} \frac{\partial}{\partial \rho}\left(\rho \frac{\partial P_3}{\partial \rho}\right) = E_3(\rho, z),$$

$$E_3(\rho, z) = V \int_0^\infty dk J_0(k\rho) \cdot \frac{\cosh(k(h-z)/\gamma)}{\sinh(kh/\gamma)} \frac{kd}{\gamma} \exp(-kd) \quad (A.1)$$

The gradient (or correlation) terms $\xi > 0$ and $\eta > 0$ (usually $\xi \approx \eta$), the expansion coefficient $\delta > 0$, while $\beta < 0$ ($\beta > 0$) for first (second) order phase transitions. The coefficient $\alpha < 0$ in the ferroelectric phase.

$$P_3(\rho \gg d, z < 0) \to -P_S, \quad \frac{\partial P_3}{\partial z}(z=0) = 0, \quad \frac{\partial P_3}{\partial z}(z=h) = 0, \quad (A.2)$$

where $P_S$ is the initial spontaneous polarization value. The boundary condition $\partial P_3/\partial z = 0$ is called "natural" [72] and corresponds to the case, when the surface energy contribution can be neglected and hence $\lambda \to \infty$ in a more general boundary condition $P_3 + \lambda(\partial P_3/\partial z) = 0$. In the case of the natural boundary conditions, a constant polarization value $P_3 = P_S$ satisfies Eq. (3b) at zero external bias, $V = 0$. For a first order ferroelectric, the spontaneous polarization in the bulk is $P_S^2 = \left(\sqrt{\beta^2 - 4\alpha\delta} - \beta\right)/2\delta$, while $P_S^2 = -\alpha/\beta$ for a second order ferroelectrics.

$$u_3(\mathbf{0}, \mathbf{y}) = \int_{-\infty}^{\infty} d_{mnk}(\mathbf{y} - \xi)\left(\int_0^h c_{jlmn} E_k(-\xi_1, -\xi_2, z)\frac{\partial}{\partial \xi_l} G_{3j}(\xi_1, \xi_2, z) dz\right) d\xi_1 d\xi_2. \quad (A.3)$$

Where $d_{ijk}(\mathbf{r}) = \frac{\partial u_{ij}(\mathbf{r})}{\partial E_k(\mathbf{r})} = 2\varepsilon_0 Q_{lmjk} \varepsilon_{il}(\mathbf{r}) P_m(\mathbf{r})$ are the stress piezoelectric tensor components ($u_{ij}$ is the strain tensor), $Q_{lmjk}$ is electrostriction tensor for cubic symmetry, $c_{kjmn}$ are stiffness tensor components. The Green's tensor $G_{ij}(\xi)$ for an elastically isotropic films is calculated in Refs.[48, 49].

Electric field could be averaged over domain volume as:

$$\langle E_3(\rho,z)\rangle = \frac{2V}{lr^2}\int_0^l dz\int_0^r \rho d\rho \int_0^\infty dk J_0(k\rho)\cdot\frac{\cosh(k(h-z)/\gamma)}{\sinh(kh/\gamma)}\frac{kd}{\gamma}\exp(-kd)$$

$$= \frac{2V}{lr^2}\int_0^\infty dk\frac{J_1(kr)}{k}d\exp(-kd)\left(1-\frac{\sinh(k(h-l)/\gamma)}{\sinh(kh/\gamma)}\right) \quad (A.4)$$

$$\underset{l=h}{=}\frac{2Vd}{r^2h}\left(\sqrt{d^2+r^2}-d\right)=\frac{2Vd}{h\left(\sqrt{d^2+r^2}+d\right)}$$

$$\langle V_c(z,\rho)\rangle = \frac{r^2h\cdot E_c}{d\left(\sqrt{d^2+r^2}-d\right)}\underset{r\sim d}{\sim} h\cdot E_c \quad (A.5)$$

When deriving Eq.(8) we substitute $\langle d_{ij}(V)\rangle$ in the expression $d_{33}^{eff}(h,d) = -\psi(h,d)(w_{333}d_{33}+w_{313}d_{31}+w_{351}d_{15})$ derived in Refs [48, 49], i.e. $d_{33}^{eff}(r) \approx -\psi(h,d)(\langle d_{33}\rangle w_{333}+\langle d_{31}\rangle w_{313}+\langle d_{15}\rangle w_{351})$.